\documentstyle[aaspp4]{article}

\begin{document}

\title{THE THERMONUCLEAR EXPLOSION OF CHANDRASEKHAR MASS WHITE DWARFS}

\author{J.C. Niemeyer and S.E. Woosley\altaffilmark{1}}
 
\affil{Max Planck Institut f\"ur Astrophysik, Karl Schwarzschild
Str.1, 85740 Garching, Germany}

\altaffiltext{1}{permanent address: Board of Studies in Astronomy and
Astrophysics, University of California, Santa Cruz, CA 95064} 

\authoraddr{Max Planck Institut f\"ur Astrophysik, Karl Schwarzschild
Str.1, 85740 Garching, Germany}

%\affil{ UCO/Lick Observatory, University of
%California, Santa Cruz, CA 95064\\
%and\\
%Max Planck Institut f\"ur Astrophysik, Karl Schwarzschild
%Str.1, 85740 Garching, Germany}

\newcommand{\ea}{{\em et al.}\,\,}

\begin{abstract}

The flame born in the deep interior of a white dwarf that becomes a
Type Ia supernova is subject to several instabilities, the combination
of which determines the observational characteristics of the
explosion. We briefly review these instabilities and discuss the
length scales for which each dominates.  Their cumulative effect is to
accelerate the speed of the flame beyond its laminar value, but
that acceleration has uncertain time and angle dependence which has
allowed numerous solutions to be proposed (e.g., deflagration, delayed
detonation, pulsational deflagration, and pulsational detonation). We
discuss the conditions necessary for each of these events and the
attendant uncertainties. A grid of critical masses for detonation in
the range $10^7$ - $2 \times 10^9$ g cm$^{-3}$ is calculated and its
sensitivity to composition explored.  The conditions for prompt
detonation are discussed. Such explosions are physically improbable
and appear unlikely on observational grounds. Simple
deflagrations require some means of boosting the flame speed beyond
what currently exists in the literature. ``Active turbulent
combustion'' and multi-point ignition are presented as two plausible
ways of doing this. A deflagration that moves at the ``Sharp-Wheeler''
speed, $0.1 g_{\rm eff} t$, is calculated in one dimension and shows
that a healthy explosion is possible in a simple deflagration if fuel
can be efficiently burned behind a front that moves with the speed of
the fastest floating bubbles generated by the non-linear
Rayleigh-Taylor instability. The relevance of the transition to the
``distributed regime'' of turbulent nuclear burning is discussed for
delayed and pulsational detonations. This happens when the flame speed
has slowed to the point that turbulence can actually penetrate the
flame thickness and may be advantageous for producing the high fuel
temperatures and gentle temperature gradients necessary for
detonation. No model emerges without difficulties, but detonation in
the distributed regime is plausible, will produce intermediate mass
elements, and warrants further study. The other two leading models,
simple deflagration and pulsational detonation, are mutually
exclusive.
\end{abstract}


\section{Introduction}
Despite 25 years of intensive investigation (e.g., Arnett 1969), the
basic physics whereby a carbon-oxygen core of nearly the Chandrasekhar
Mass (1.39 M$_{\odot}$) explodes as a Type Ia supernova (SN Ia) is still
debated. One may reasonably conclude that it is a hard problem. In
fact, only recently has the astrophysics community begun to profit
from the extensive experience of the chemical combustion community in
order to appreciate fully just how complicated burning coupled to
hydrodynamics really can be.

The astrophysical problem is especially hard because the nuclear flame
propagates in an extensive medium in which gravity plays a role and
several instabilities have time to develop over a large range of
length scales. Any realistic solution must take cognizance of these
instabilities and, if only by parameterization, incorporate them into
the stellar model.

This need, and variability in the outcome depending on uncertain
parameters, has given rise to several classes of supernova models, all
largely empirical. These include the ``delayed detonation" (Khokhlov
1991abc; Woosley \& Weaver 1994), ``pulsational detonation" (Arnett \&
Livne 1994ab); ``pulsational deflagration" (Nomoto, Sugimoto, \& Neo
1976; Ivanova, Imshennik, \& Chechetkin 1974); ``convective
deflagration" (Nomoto, Thielemann, \& Yokoi 1984); and the fractal
model of Woosley (1990), each of which, by various contrivances,
generates a flame which is born slow and accelerates very rapidly as
the star begins to come apart. This behavior has been found essential
to obtaining nucleosynthesis that agrees with spectroscopic
observations.

There are two simple solutions to the explosion problem, neither of
which is thought to be correct, but which bound the real solution - a)
a laminar conductive flame, and b) prompt detonation. The latter is
improbable (\ref{prompt}) and would give unacceptable
nucleosynthesis - no intermediate mass elements; the former is
unphysical and would not give an energetic explosion.  Between these
two extremes lies the regime of unstable flame propagation to which
this paper is mostly devoted. We begin by reviewing the relevant
instabilities and current views regarding their importance and mutual
couplings. Though much has been written, especially regarding the
Rayleigh-Taylor (RT) instability, not all share the same views even on
this fundamental 
subject. It is therefore necessary to state (our view of) the basics
before proceeding to the models, and so we briefly discuss (\ref{dyn})
the RT-instability, the Landau-Darrieus (LD) instability, the
Kelvin-Helmholtz (KH) instability, and turbulence. Our main goals here are:
1) ascertaining the uncertainties in the models in an attempt to
resolve the leading candidate(s); and 2) exploring the implications of
new physics - active turbulent combustion, multipoint ignition, and,
especially, detonation in the distributed regime - for models for SN Ia's.

It is frequently stated (Khokhlov 1995, Arnett \&
Livne 1994a) that simple deflagrations, those in which the flame remains at
all times subsonic and in which an intervening pulse of the white
dwarf does not occur, cannot give an energetic explosion. We discuss
why this may not be true, either because of physics that has been left
out of the calculations (\ref{act}, \ref{sync}) or because the flame
has not been modeled sufficiently accurately in three dimensions
(\ref{simdefl}).

Another important point that has not been adequately discussed in the
context of astrophysical flames, is the idea of ``distributed
burning''. So long as the thickness of the flame is negligible, it can
be treated as a temperature discontinuity. Heat never moves 
far from the burning surface. This is the case during most of the
flame's life. But for densities below $\sim 3 \times 10^7$ g cm$^{-3}$
turbulence will disrupt the flame sheet. Once this happens, one can no
longer speak of the flame as either laminar or conductive.  It has
entered a domain where turbulent transport directly affects the
nuclear reaction region and smears it out over macroscopic length
scales. 
%heat can be extracted, by turbulence, both from
%the flame itself and from the ashes, and used directly to heat the
%unburned carbon. 
The combustion community calls this type of burning
the ``distributed flame regime" (e.g., Pope 1987) and it is a frontier
topic for them as well as us.  While the increased mixing of hot ash
and cold fuel can, in
some instances, be beneficial for provoking a detonation, the same
process also leads to compositional mixing which increases the
critical mass required for detonation to occur (\ref{mcrit}). In
section (\ref{dist}) we discuss the physical conditions required for
this sort of 
detonation and show that it is a reasonable, if uncertain,
occurrence. In this sense, it is superior to many other models in the
literature.

Much has been written about other forms of the ``delayed detonation
model'' (e.g., Khokhlov 1991abc; Woosley \& Weaver 1994), yet the
physics of the transition to detonation remains obscure. In sections
(\ref{dd1}) and (\ref{dist}), we discuss the two different kinds of delayed
detonation that have been previously published and why each is
unlikely to occur.

In the conclusions (\ref{conc}) we summarize our results.

\section{FLAME DYNAMICS AT HIGH DENSITIES}
\label{dyn}
\subsection{The conductive laminar flame}
\label{cond}

The simplest solution to the propagation of burning in a premixed fuel is an
elementary ``flame". Heat is transported ahead of the burning region, in this
case, by electron conduction. The temperature rises to the point where
reactions can consume the fuel (carbon) on a diffusive
time scale and this condition sets both the thickness of the flame and its
steady velocity (Landau \& Lifshitz 1991). In the absence of instabilities,
these quantities can be determined analytically with considerable precision
(Timmes \& Woosley 1992).

The critical mass
required to keep a flame alive is small, a few times
$4/3 \pi \rho l_{th}^3$, with $l_{\rm th}$, the flame thickness. For
isolated regions below this mass, heat
can diffuse out and the flame will die, but so long as a critical mass
remains intact, one cannot extract heat from the ashes of the
combustion (in order to raise the temperature of the fuel) over a
greater distance than $l_{\rm th}$. It is impossible for a simple
{\sl laminar} flame to turn into a detonation. So long as there
is a flame with a well defined surface (exceptions to this will be
discussed in sections \ref{sum} and \ref{dist}), detonation can only
be achieved by increasing the area of the flame.

It is well known that the laminar speed is too slow to make a
supernova with the observed properties. In fact, we shall conclude
(see also Niemeyer \& Hillebrandt 1995b and Khokhlov 1995) that the
laminar speed is not very relevant for the effective rate at which
burning spreads.  Turbulence and Rayleigh-Taylor instabilities are
more important and carry the flame at a speed independent of the
microphysics. However, it is important to keep track of the flame {\sl
thickness}, as this may ultimately affect even the macroscopic nature
of the burning, and to distinguish the speed with which the outer
boundary of burning spreads from the total rate of mass
consumption. These rates are quite different if a large amount of fuel
becomes entrained.

After a brief review of the various instabilities that arise from a
linear analysis of flame propagation under the influence
of gravity and shear in section (\ref{inst}), we discuss the
quasi-stationary structure of the flame surface after it reaches the
fully nonlinear regime in sections  (\ref{cells}), (\ref{tur}) and (\ref{rt}).

\subsection{Linear instabilities}
\label{inst}

The problem of linear hydrodynamical stability of subsonic
flames in the thin flame representation was first analyzed by Darrieus
(1938) and, independently, by Landau (1944). Landau's result for the
linear growth rate also
includes the influence of gravity on the density jump produced by the
flame front, which is equivalent to the Rayleigh-Taylor instability
(Chandrasekhar 1961) if the flame speed is set to zero. In the context
of nuclear flames in degenerate 
carbon, the LD instability has been
explored in detail both analytically and numerically by Blinnikov \& Sasorov
(1996) and Bychkov \& Liberman (1995), and its onset has been
demonstrated by means of two 
dimensional hydrodynamical simulations (Niemeyer \& Hillebrandt
1995a). All these studies agree that 
on scales larger than the Markstein, or critical, length $l_{\rm crit}
\approx 100\,l_{\rm th}$ (Markstein 1951), flames moving upward are
unstable to both LD and RT instabilities on all wavelengths.

In addition to radial perturbations, we have to consider perturbations that
face in a direction perpendicular to gravity, where buoyancy of burned
material floating in the cold fuel induces tangential velocity
differences along the flame surface. Here, another instability becomes
important, the KH or shear instability (e.g., Landau
\& Lifshitz 1991). In our context, it is quite
important to know the circumstances under which the flame surface in a Type
Ia supernova is KH unstable since this condition marks the transition to the
fully turbulent burning regime. Strictly speaking,
tangential discontinuities can only occur if the mass flux through the
surface vanishes, for otherwise continuity of the momentum flux imposes
continuity of the tangential velocity components. The mass flux
through the burning zone is by definition
non-zero. Thus, we can qualitatively argue that the propagation of a
flame tends to stabilize the front against KH-instability. If, however, the
flow field around a burning bubble is dominated by buoyant
acceleration, i.e.~the mass flux becomes small compared with
the velocity 
components tangential to the front, the flame becomes KH unstable, as
shown by numerical simulations (Niemeyer
\& Hillebrandt 1996). This occurs during the nonlinear stage of the
RT-instability. 

\subsection{The cellular regime}
\label{cells}

In the nonlinear evolution of the LD instability, the
formation of cusps at the points where the flame front
self-intersects  gives rise to an additional 
quadratic damping term for the perturbation amplitude that is not
included in Landau's linear stability analysis (Zeldovich \ea
1985). It stabilizes the flame surface after cells with a
stationary, scale independent amplitude have formed. The speed of the emerging
cellular surface is directly given by the increased surface area,
yielding
\begin{equation}
\label{1cell}
u_{\rm cell} = u_{\rm lam}(1 + \epsilon(\mu))\,\,,
\end{equation} 
where the velocity increment $\epsilon$ is a function of the gas
expansion parameter $\mu = \rho_{\rm b}/ \rho_{\rm u}$ (Zeldovich 1966). 
If $\mu$ is small, as in the case of burning in degenerate matter, 
$\epsilon$ can be approximated by (Zeldovich 1966)
\begin{equation}
\label{epszel}
\epsilon(\mu) = \frac{\pi^2}{24}\,(1 - \mu)^2 \approx 0.41 (1 - \mu)^2 \,\,.
\end{equation}

Motivated by the large dynamical range of thermonuclear flames in
white dwarfs Blinnikov and Sasorov (1996) proposed a fractal
model for the cellular structure of LD unstable flames. By means of
a statistical analysis of the Sivashinsky equation for thin flame
propagation, the authors estimated the fractal dimension of one
dimensional flames as
\begin{equation}
D_{\rm 1d} = 1 + D_0 \gamma^2
\end{equation}
where $\gamma = (1- \mu)$. This result was 
confirmed by numerical simulations of the closely related Frankel equation,
which yielded $D_0 \approx 0.3$. Furthermore,
the authors derived the fractal dimension of two-dimensional flame
surfaces as $D_{\rm 2d} \approx 2 D_{\rm 1d}$.

Here, we take a simplified approach that shows how $D_0$ of
one-dimensional cellular flame fronts can be
related to $\epsilon$. By doing so, we neglect some subtle, but
important, complications that arise in the statistical treatment of
the Sivashinsky equation which are accounted for in the more
sophisticated approach of Blinnikov and Sasorov (1996). In our simple
model, we describe the cellular 
front as a hierarchy of cells on all length scales. If consecutive
cell-generations are widely separated, we can assume
that the thin flame
approximation is valid on each scale. The effective propagation speed $u_i$
on scale $l_i$ is then related to scale  $l_{i-1} < l_i$ by
\begin{equation}
u_i = u_{i-1} +  \epsilon u_{i-1} \,\,.
\end{equation}
Taking the continuum limit and integrating yields
\begin{equation}
u_i = u_0 e^{\epsilon i}\,\,.
\end{equation}
We now assume that cell splitting occurs after a dilation interval
$S$, so that $l_i = S l_{i-1}$. If the smallest unstable
length scale is of the order of the Markstein length, $l_0 = l_{\rm
crit}$, we find $l_i = S^i l_{\rm crit}$ and we can express the 
effective flame velocity in terms of the length scale $l$:
\begin{equation}
\label{ucell}
u_{\rm cell}(l) \approx u_{\rm lam} \left(\frac{l}{l_{\rm
crit}}\right)^{\epsilon/{\rm ln}S}\,\,.
\end{equation}
If the flame speed is interpreted in 
a geometrical way, i.e., $u(l) \propto \bar A_l$, where $\bar A_l$ denotes
the increased surface of a cellular front observed at the scale $l$, it follows
that the surface area behaves like a fractal (Mandelbrot 1983)
\begin{equation}
\label{cellsurf}
\bar A_l = A_{\rm lam} \,\left(\frac{l}{l_{\rm crit}}\right)^{D_{\rm
cell} - 1}
\end{equation}
with the fractal dimension
\begin{equation}
\label{Dcell}
D_{\rm cell} = 1 + \frac{\epsilon}{{\rm ln}S}\,\,.
\end{equation}

Inserting (\ref{epszel}) into (\ref{Dcell}) we find, in agreement
with Blinnikov and Sasorov (1996), that the fractal excess of (\ref{Dcell})
is proportional to $\gamma^2 = (1- \mu)^2$. Specifically, we find $D_0
= \pi^2/24\,{\rm ln}S$, and agreement with the authors'
numerical results yields a dilation interval, $S \approx 4$.  

\subsection{Nuclear burning in the flamelet regime}
\label{tur}

In the case of a cellular flame front driven purely by the
LD instability, i.e. in the absence of gravity, there is no known
upper limit for the largest scale of cell formation. However, if
gravity is turned on the cell structure is no longer scale independent,
which leads to 
the break-down of nonlinear stabilization (Khokhlov 1995).
As soon as the process of cell disruption and bubble formation occurs
on the largest scales, a cascade of turbulent velocity fluctuations is
produced that continues on scales below the actual large scale flame
instability (Niemeyer \& Hillebrandt 1995b). After the establishment
of the turbulent cascade, there exists a range of scales
where burning is dominated by isotropic, 
fully developed turbulence. We will restrict our discussion to the
conservative assumption 
that the production of turbulence is provided purely by large
scale fluid instabilities, so that turbulent burning can be called
``passive''. Some thoughts on ``active'' turbulent 
combustion, where thermal expansion within the burning region is
assumed to influence the properties of turbulence, are given in
section (\ref{act}).

Cell formation ceases on the scale $l$ where the turbulent velocity
fluctuations $v(l)$ become comparable with  $u_{\rm
cell}(l)$. Here, the time of front interaction with turbulent
eddies becomes comparable to the eddy turnover time and, consequently,
perturbations caused by turbulence grow to amplitudes comparable
with the cell amplitudes. The cellular flame front is thus unstable on these
scales (Zeldovich \ea 1980). Furthermore, numerical simulations of
curved flames subject to shear show the breakdown of the nonlinear
stabilization mechanism if $v_{\rm shear} \approx u_{\rm cell}$
(Niemeyer \& Hillebrandt 1996). The transition between
cellular and turbulent burning regimes
is therefore marked by the Gibson scale, defined as (Peters 1988):
\begin{equation}
\label{lgibs3}
v(l_{\rm gibs}) = u_{\rm cell}(l_{\rm gibs})\,\,,
\end{equation}
Note that, due to the arguments above, we take
eq.~(\ref{lgibs3}) as an equality, which differs from earlier results
(Khokhlov 1995) where the smallest turbulent scale was found to be
$\approx 500\, l_{\rm gibs}$. Using
the Kolmogorov scaling law  $v(l) \propto l^{1/3}$ it
follows that $l_{\rm gibs}$ 
scales with the third power of $u_{\rm cell}/v(L)$, where $v(L)$ is,
for instance, the magnitude of the turbulent velocity on the largest
turbulent scale $L$. If we assume that near the
beginning of the explosion  this ratio is close to unity, implying
that $l_{\rm gibs} \approx L$, and later decreases owing
to the decreasing flame speed and the build-up of shear in the
RT mixing region, we find that the Gibson
length decreases continually during the explosion. An upper bound for
the intensity of turbulence on large scales is given by the ``freezing
out'' of turbulent motions as a result of the overall expansion of the
star (Khokhlov 1995). Using $v(L) \lesssim 10^7$ cm s$^{-1}$ at $L
\approx 10^6$ cm and $u_{\rm cell} \approx u_{\rm
lam} \gtrsim 10^5$ cm s$^{-1}$ (Timmes \& Woosley 1992), we estimate that
$l_{\rm gibs} \gtrsim 1$ cm. Consequently,
$l_{\rm gibs} \gtrsim l_{\rm th}$ for all densities $\rho \gtrsim 3 \times
10^7$ g cm$^{-3}$. For this reason, turbulent nuclear flame fronts
in white dwarf matter at these densities burn in the  ``corrugated
flamelet regime'' (e.g., 
Clavin 1994). At lower densities, the flame front enters the so-called
``distributed regime'' that will be described in sections (\ref{sum})
and (\ref{dist}). 

The flamelet regime is characterized by laminar flame propagation on
microscopic scales, while burning is determined purely by turbulence on
large scales and therefore independent of microphysics.
Fuel ``digestion'' occurs in an
extended region behind the boundary of fuel and ashes, the so-called
``turbulent flame brush''. The size of 
this region and the range of turbulent length scales adapt in a way
that provides a flame brush propagation velocity that is
decoupled from the laminar flame speed. On scales
obeying $l \gg l_{\rm gibs}$, one can express the effective flame 
speed $u_{\rm tur}(l)$ and turbulent front width, $d_{\rm tur}(l)$ in
terms of $l$ and the turbulent velocity $v(l)$ (Kerstein 1988, Clavin
1990), since both transport of burning fluid
into the fresh material and fuel
consumption inside the flame brush are limited by the eddy turnover time
$\tau_{\rm t}(l) \approx l/v(l) \propto l^{2/3}$.
Consequently, the relations
\begin{equation}
\label{utur}
u_{\rm tur}(l) \approx v(l) \quad , \quad d_{\rm
tur}(l) \approx l
\end{equation}
are reasonable order of magnitude approximations that have been
employed extensively in combustion research in the limit $v(l) \gg
u_{\rm lam}$.

\subsection{The Rayleigh-Taylor regime}
\label{rt}

On the largest scales ($l \gtrsim 10^6$ cm) the flame dynamics is
dominated by buoyancy of the hot, burned material surrounded by
denser carbon and oxygen. The linearized problem is expressed by
the RT instability (Chandrasekhar 1961). After a
short period of exponential growth, the perturbation amplitudes become
comparable to their wavelengths and the structure enters the nonlinear
stage where interactions among the structures can
occur, giving rise to merging and fragmentation of bubbles. Another
important process is the appearance of 
KH unstable regions along the bubble surfaces. These produce turbulent eddies
that spread out bubble tips and walls.
Finally, the long time evolution shows continuing merging and fragmentation of
rising bubbles, creating an increasingly turbulent mixing layer
(Snider \& Andrews 1994).

Experiments of gas bubbles rising in vertical tubes filled with fluid
were performed to measure the asymptotic velocity of a single bubble
(Davies \& Taylor 1949). Best fits to the measurements were obtained
by the relation
\begin{equation}
\label{vrt}
v_{\rm rt} = {\cal B}\,\sqrt{g_{\rm eff} l}\,\,,
\end{equation}
where ${\cal B} \in [0.466,0.490]$ is a constant, $g_{\rm eff} = At\, g$
is the effective gravitational acceleration ($At = (\rho_{\rm u} - \rho_{\rm
b})/(\rho_{\rm u} + \rho_{\rm b}) \approx 0.5 (1 - \mu)$ is
the Atwood number), 
and $l$, the radius of the tube. Layzer (1955) solved the
problem analytically for a spherically symmetric tube and derived ${\cal B}
= 0.511$. Khokhlov (1995) derived a similar prescription for the
propagation speed of RT unstable flames in open boxes from numerical
simulations. While this result is true for a single
length scale, we need to consider a range of RT-unstable scales, where
the front creates a large number of bubbles
with various radii. The so-called ``Sharp-Wheeler Model''
statistically describes the nonlinear stage of a multiscale RT-front (Sharp
1984). It consists of a one-dimensional string of bubbles that are
described by their radius, $r_i$, and height, $z_i$. According to
(\ref{vrt}), the heights grow as $\dot z_i = v_{\rm
rt}$. Neighboring bubbles merge if the difference between their
heights exceeds the radius of the smaller bubble and thus the average bubble
radius increases with time. Consequently, the average rise
velocity from eq.~(\ref{vrt}) also grows. Numerical simulations of the
Sharp-Wheeler-model (Glimm \& Li 1988) show that the
front asymptotically attains a constant acceleration that is
proportional to $g_{\rm eff}$, so that the RT-mixing region grows at a
rate of approximately 
\begin{equation}
\label{rsw}
r_{\rm sw} \approx 0.05 g_{\rm eff}t^2\,\,.
\end{equation}
This result is in
agreement with full hydrodynamical simulations (Young 1984) and
experiments (Read 1984). Equivalently, we can say that the front
advances into the cold material with a speed of 
\begin{equation}
\label{vsw}
v_{\rm sw} \approx 0.1 g_{\rm eff}t\,\,.
\end{equation}
Comparing (\ref{vrt}), (\ref{rsw}) and (\ref{vsw}) shows that the
maximum bubble
radius evolves linearly with its displacement from the stellar center
(i.e., the inner boundary of the RT mixing zone), $l_{\rm max} \propto
r$.

In the context of supernova modeling, we are mainly interested in the
burning velocity of the RT unstable flame brush. As in section
(\ref{tur}), we can 
argue that the overall burning rate is limited by the {\it fastest}
transport mechanism that mixes ashes and fuel. Invoking the arguments
of the previous section, fuel consumption automatically adjusts to the speed 
of fuel contamination by burning blobs in order to provide a burning
rate that is independent of microphysics. The highest speed for each
single length scale $l$ is now given by (\ref{vrt}), yielding $u_{\rm rt}
\propto l^{1/2}$.
A geometrical interpretation (eq.~\ref{cellsurf}) of the flame
speed in the nonlinear RT-regime therefore yields a fractal dimension
of the flame surface, $D_{\rm rt} = 2.5$. In addition, we need the
evolution the maximum bubble size, $l_{\rm max}$, as a function
of time or radial displacement, like the one provided by the
Sharp-Wheeler model. If we assume that bubble growth is purely
governed by merging, and that burning is completed somewhere within the
RT mixing region (this need not necessarily be the case), the effective
burning velocity is given by the Sharp-Wheeler speed (\ref{vsw}).
One dimensional supernova models using eq.~(\ref{vsw}) will be presented
in section (\ref{simdefl}).

Finally, we define the boundary between the turbulent burning regime
dominated by Kolmogorov scaling and the buoyancy driven
RT-regime by looking at the minimum time scale for self-interaction of flame
structures with the size $l$, roughly given by $\tau_{\rm si}(l) \approx
l/v_{\rm rt}(l) \propto l^{1/2}$. As stated above, expansion of the
star inhibits the
growth of the largest structures into the fully nonlinear regime where they
become isotropically turbulent. Expansion is characterized by the
hydrodynamical time scale $\tau_{\rm dyn} \approx 0.1$ s, so that
self-interaction resulting in fully developed turbulence occurs on
scales below $l_{\rm tur/rt}$ defined by $\tau_{\rm si}(l_{\rm
tur/rt}) = \tau_{\rm dyn}$. Inserting equation (\ref{vrt}) yields
\begin{equation}
\label{lrt}
l_{\rm tur/rt} = \tau_{\rm dyn}^2\,{\cal B}^2\,g_{\rm eff} \approx
10^{6}\,{\rm cm}\,\,,
\end{equation}
where we have used $g_{\rm eff} \approx 5 \times 10^8$ cm s$^{-2}$ as a
typical value for the effective gravitational acceleration.

\subsection{Summary of instabilities and their effects}
\label{sum}

\begin{figure}
%\epsscale{0.8}
\plotone{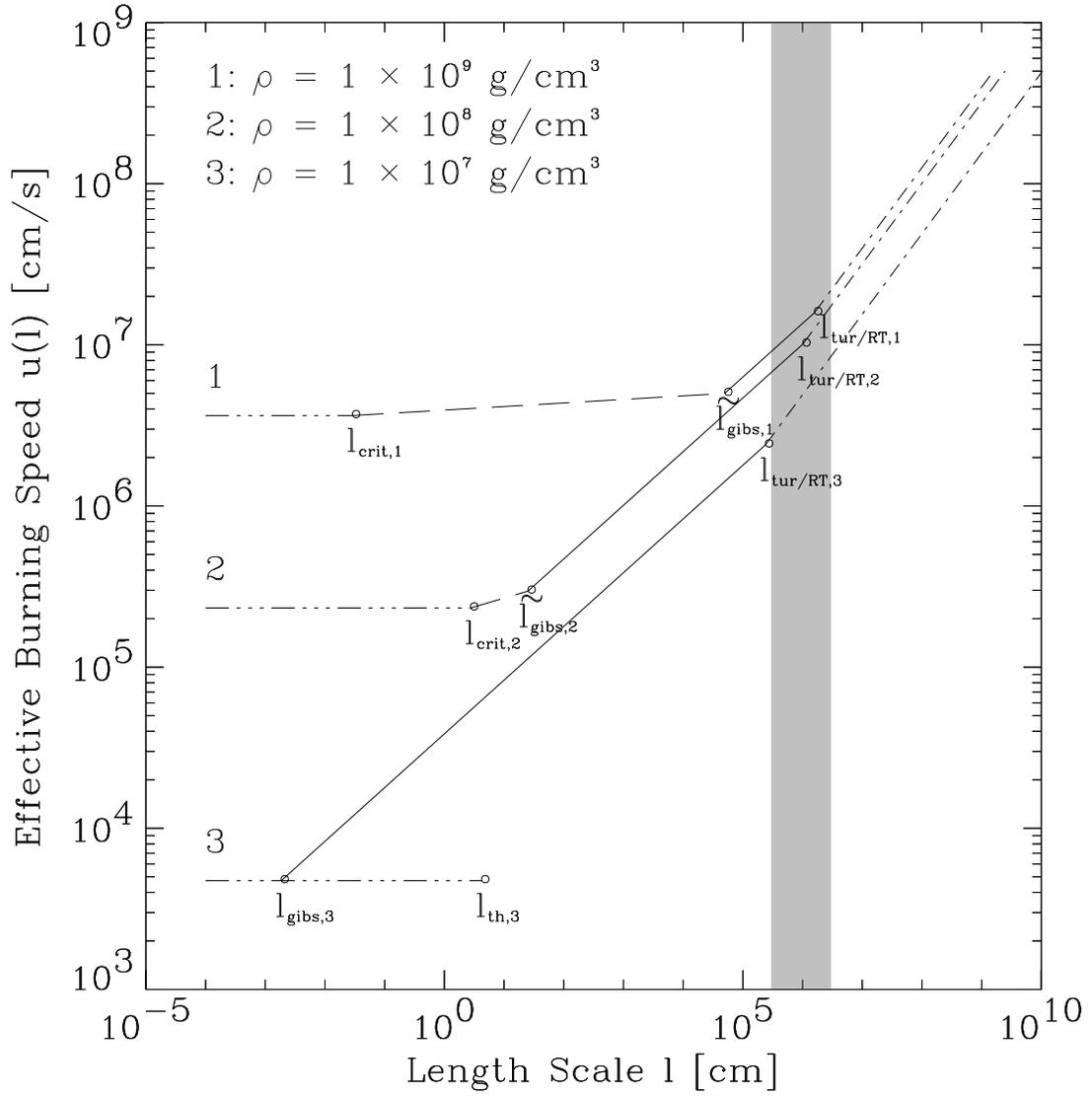}
\caption{\label{ul} Effective burning speed as a function of length 
scale for the early, main, and late phase of the deflagration. The
shaded region represents the typical resolution of multidimensional
simulations, $\Delta \approx 10$ km.}
\end{figure}

Beginning at the smallest dynamically relevant scale, the
thermal flame thickness, $l_{\rm th}$, the flame propagates with the
laminar flame speed, $u_{\rm lam}$, until the smallest cells appear at
$l_{\rm crit} \approx 100\, l_{\rm th}$. Therefore,
$u(l) = {\rm
const}(l)$. In the cellular regime, $u(l) \propto l^{D_{\rm cell}-2} \approx
l^{0.1}$. Cellular stabilization fails
when turbulent velocities become comparable with the effective cellular
flame speed at $l_{\rm gibs}$. As a result of the Kolmogorov
scaling law for turbulent velocity
fluctuations and the assumption that the turbulent flame speed is
determined by the eddy turnover time on every scale, we find that
$u(l) \propto l^{1/3}$ in the fully turbulent regime. Above  $l_{\rm
tur/rt}$, the largest upward velocity on scale 
$l$ is determined by buoyancy. Hence, $u(l) \propto l^{1/2}$ according
to (\ref{vrt}). 
Fig.~(\ref{ul}) is a summary of the scale dependence of the
burning speed $u(l)$. The piecewise scale-invariant burning regimes
are represented by straight lines with different slopes in the
log-log-plot. Three separate curves are
shown, corresponding to the characteristic velocities and length
scales at early, central, and late times of the deflagration phase (as
represented by three decreasing density values). All graphs are
based on the assumption that all the relevant instabilities and
the transitions between them have reached a statistical equilibrium
state. This is true after approximately one growth time of the largest
considered scale ($\approx 0.2$ s).

The curves in Fig.~(\ref{ul}) are constructed as follows: First, the
largest fully turbulent length scale  $l_{\rm tur/rt}$ is computed
from equation (\ref{lrt}) with values for $g$ taken from Khokhlov
(1993a), while the burning speed at this scale is
determined from equation (\ref{vrt}). This point in
the $u$-$l$-plane
serves as the origin for the purely buoyant part with $u(l) \propto
l^{1/2}$ extending to larger $l$, and for the Kolmogorov part reaching
downward. The laminar flame speed is used as the second absolute point
of each graph. Its value, as well as the thermal flame thickness,
$l_{\rm th}$, and the expansion factor, $\gamma = \Delta \rho / \rho$,
are taken from Timmes \& Woosley (1992). At a length of $l_{\rm crit}
\approx 100\,l_{\rm th}$, the LD-instability and the subsequent
formation of cells mark the transition to the scaling $u(l) \propto
l^{D_{\rm cell}-2}$ with $D_{\rm cell} \approx 2(1+0.3\gamma^2)$
(Blinnikov \& Sasorov 1996; \ref{cells}). This line is extended until it
intersects the Kolmogorov line coming from above, which defines the
Gibson scale $l_{\rm gibs}$. Notice that, by virtue of
this construction procedure reflecting our current understanding of
flame dynamics in scale space, all burning properties above $l_{\rm
gibs}$ are uniquely fixed by large scale phenomena and 
therefore independent of microphysics.

The early phase of the explosion, where $\rho \approx 10^9$ g cm$^{-3}$, is
characterized by a high laminar flame speed and small thermal
expansion, resulting in a very shallow slope of $u(l)$ in the cellular
regime. On scales of approximately 10 km, however, both effects are
already dominated by the RT-instability and turbulent burning. The
turbulent regime becomes more pronounced as $u_{\rm lam}$ declines and
the Gibson scale decreases ($\rho \approx 10^8$ g cm$^{-3}$, second
graph). Meanwhile, the cellular part of $u(l)$ becomes steeper owing
to the increasing thermal expansion $\gamma$. In the simplified
construction method outlined above, the highest turbulent velocities
are assumed to be coupled directly to the RT-velocity at the
transition length  $l_{\rm tur/rt}$. Consequently, the model implied
here ``forgets'' about all turbulence that has been built up by
earlier RT-fluctuations, in contrast with the more realistic
expectation that the level of turbulence decreases according to the
almost negligible microscopic dissipation into heat. This
assumption of instantaneous adjustment results in decreasing
RT- and turbulent flame speeds as the density -- and therefore
$g_{\rm eff}$ in equation (\ref{lrt}) -- falls. A phenomenological
approach to account for the memory effect of accumulated turbulence in
numerical simulations of SN Ia's has been proposed by Niemeyer \& Hillebrandt
(1995b).

\begin{figure}
\plotone{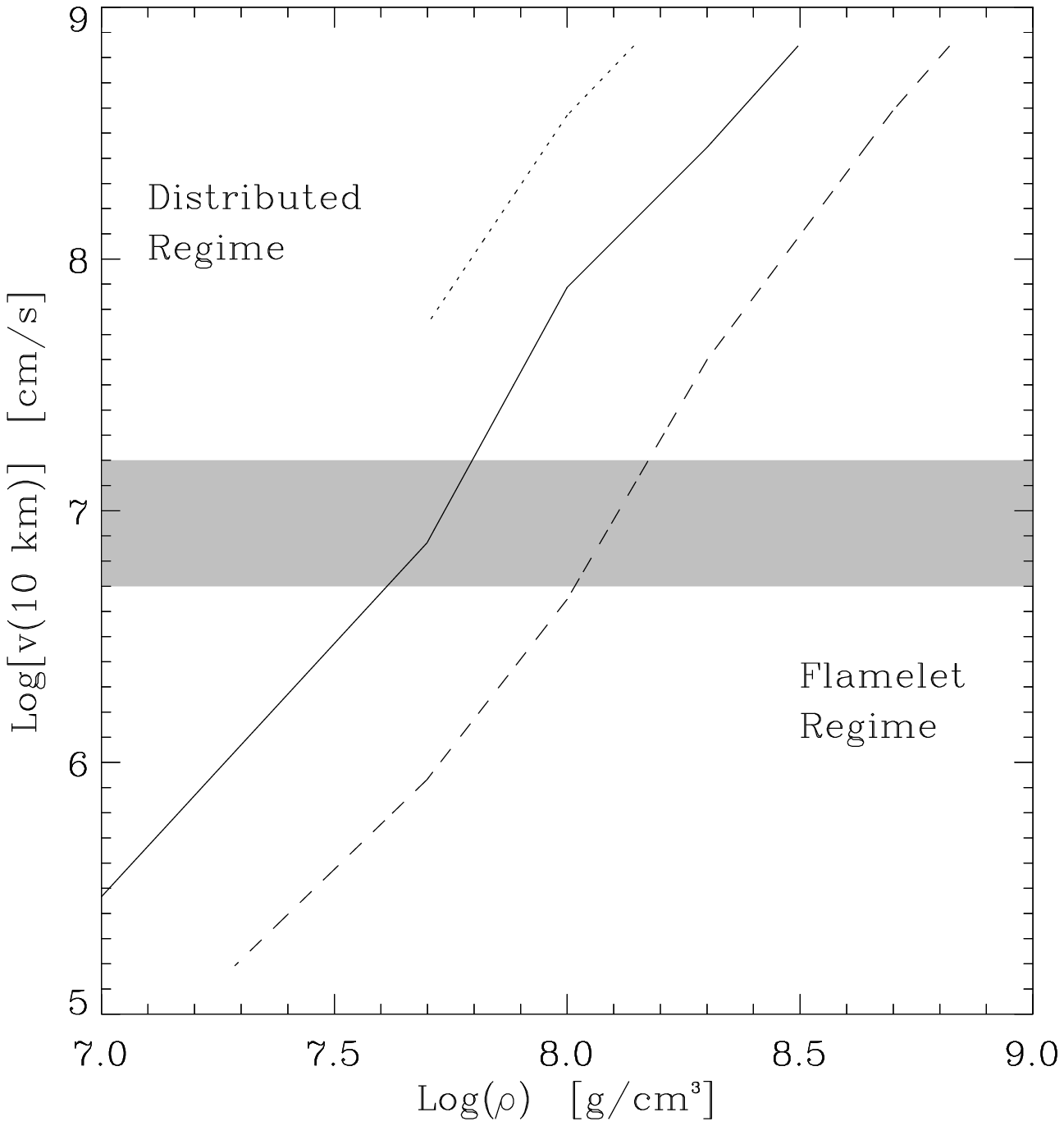}
\caption{\label{trans} Regions of flamelet and distributed burning 
depending on fuel density and turbulent velocities at $L \approx 10$
km. The lines depict $l_{\rm th} = l_{\rm gibs}$ for different
compositions (- - - $X_{\rm C} = 0.2$, $\cdot \cdot \cdot X_{\rm C} = 1$,
--- $X_{\rm C} =  0.5$). The shaded region shows the range of
turbulent fluctuations induced by the RT-instability on large scales.}
\end{figure}
Fig.~(\ref{ul}) also shows that our argumentation breaks
down at the lowest density ($\rho \approx 10^7$ g cm$^{-3}$), represented
in the bottom graph. Here, the Gibson length, $l_{\rm gibs}$ is smaller
than the thermal flame thickness, $l_{\rm th}$, by 
almost four orders of magnitude. Therefore, the conditions for the
flamelet regime (section \ref{tur}) are no longer satisfied, implying
that laminar flame propagation ceases to occur on all scales (strictly
speaking, this also implies that $l_{\rm gibs}$ is no longer a
well defined quantity). Instead,
turbulent eddies disrupt the burning region and dominate over
conductive transport even on microscopic scales. This burning
regime is sometimes termed ``distributed'' or ``stirred combustion''
(e.g., Pope 1987). We emphasize that all
of the preceding discussion is based sensitively upon the assumption of
microscopic flamelets and therefore becomes largely invalid after the
transition to the distributed burning regime. Regions of distributed
and flamelet regimes are displayed in Fig.~(\ref{trans}) as a
function of fuel density and macroscopic turbulent velocity
$v(10$km). The lines depict the transition defined by $l_{\rm th} =
l_{\rm gibs}$ for three different compositions. Decreasing carbon mass
fraction is related to an earlier transition to distributed burning,
since the flame thickness grows while the flame speed decreases.

\section{HOW THE WHITE DWARF EXPLODES}

\subsection{Deflagration}
\label{defl}

\subsubsection{A simple deflagration?}
\label{simdefl}

The simplest outcome to the exploding white dwarf problem would be a
supernova in which the flame remained at all times and places subsonic
and burned a sufficient fraction of the stellar mass to explode the
star violently on the first attempt with no intervening
pulsation. This is the basis for the common ``deflagration" or
``convective deflagration" model of which W7 (Nomoto, Thielemann, and
Yokoi 1984) is a popular example.  Numerous studies of this model (and
others like it, e.g., Woosley, Axelrod, \& Weaver 1984; Woosley \&
Weaver 1994) give at least moderately good agreement with the observed
light curve, spectrum, and nucleosynthetic requirements of common Type
Ia events. The model does so by virtue of having approximately the correct
proportions of intermediate mass elements (Si-Ca) to $^{56}$Ni and the
right final $^{56}$Ni mass for a Chandrasekhar mass starting
point. This result is, in turn, a consequence of a flame speed that
has been crafted to yield the desired result by a particular choice of
transport algorithm (mixing length convection) and scale parameter
(the mixing length as a function of time). Knowing approximately the
desired result is useful, but can the flame speed in a real
deflagration behave in the required way?

Recent multi-dimensional calculations differ in their conclusions
regarding this important issue. Khokhlov (1995) and Arnett \& Livne
(1994a) find that the flame moves too slowly to explode the star on
the first try. Niemeyer \& Hillebrandt (1995b) and Niemeyer \ea (1996) find
that a prompt explosion is possible, albeit a weak one that makes
little $^{56}$Ni. A key difference in these calculations is the
treatment of turbulence. All assume that the RT-instability is chiefly
responsible for the production of turbulence, but while the former
models assume that RT structures are isotropic and therefore
implicitly employ an instantaneous adjustment of the turbulent cascade
to the RT fluctuations, the latter simulations allow for a time delay
between the production and dissipation of turbulent kinetic
energy. They therefore include a ``memory'' for the history of the
turbulent energy of each fluid element which, in some cases, gives
rise to a build-up of turbulence above its local equilibrium.

In the next sections, two physical ways of boosting the deflagration
efficiency are discussed. A third possibility is simply that
calculations having the necessary three dimensional resolution and low
numerical viscosity have yet to be done. Were it not for drag
(admittedly an unreasonable omission), a buoyant bubble floating in
response to its density contrast ($\sim$40\% typically when the flame
is at 1000 km) and local gravity ($\sim 10^{10}$ cm s$^{-2}$) would
quickly achieve the sound speed. Assuming that the bubble dimension
scaled linearly with distance from the center of the star, a
reasonable assumption in the bubble cascade model of Sharp and Wheeler
(Sharp 1984; \ref{rt}), specifically taking the bubble radius, $l_{\rm max}
\sim 0.5 r$ with $r$ the distance to the stellar center and using
eq.~(\ref{vrt}), one obtains effective flame speeds $\sim$1500 km
s$^{-1}$. This prescription implicitly includes drag and the result is
consistent with what is seen by Garcia-Senz \& Woosley (1995) in
analytic calculations and by Niemeyer \ea (1996) in two-dimensional
numerical models. A similar result can be obtained by directly using
the Sharp-Wheeler burning velocity (\ref{vsw})  which, for times 0.5 to
1 sec, is again of order 1000 to 2000 
km s$^{-1}$. Provided turbulence is capable of burning out or
``digesting'' all material internal to the leading bubbles, this speed
would be enough to generate at least a mild explosion (consistent with Niemeyer
\ea 1996).

\begin{figure}
\plotone{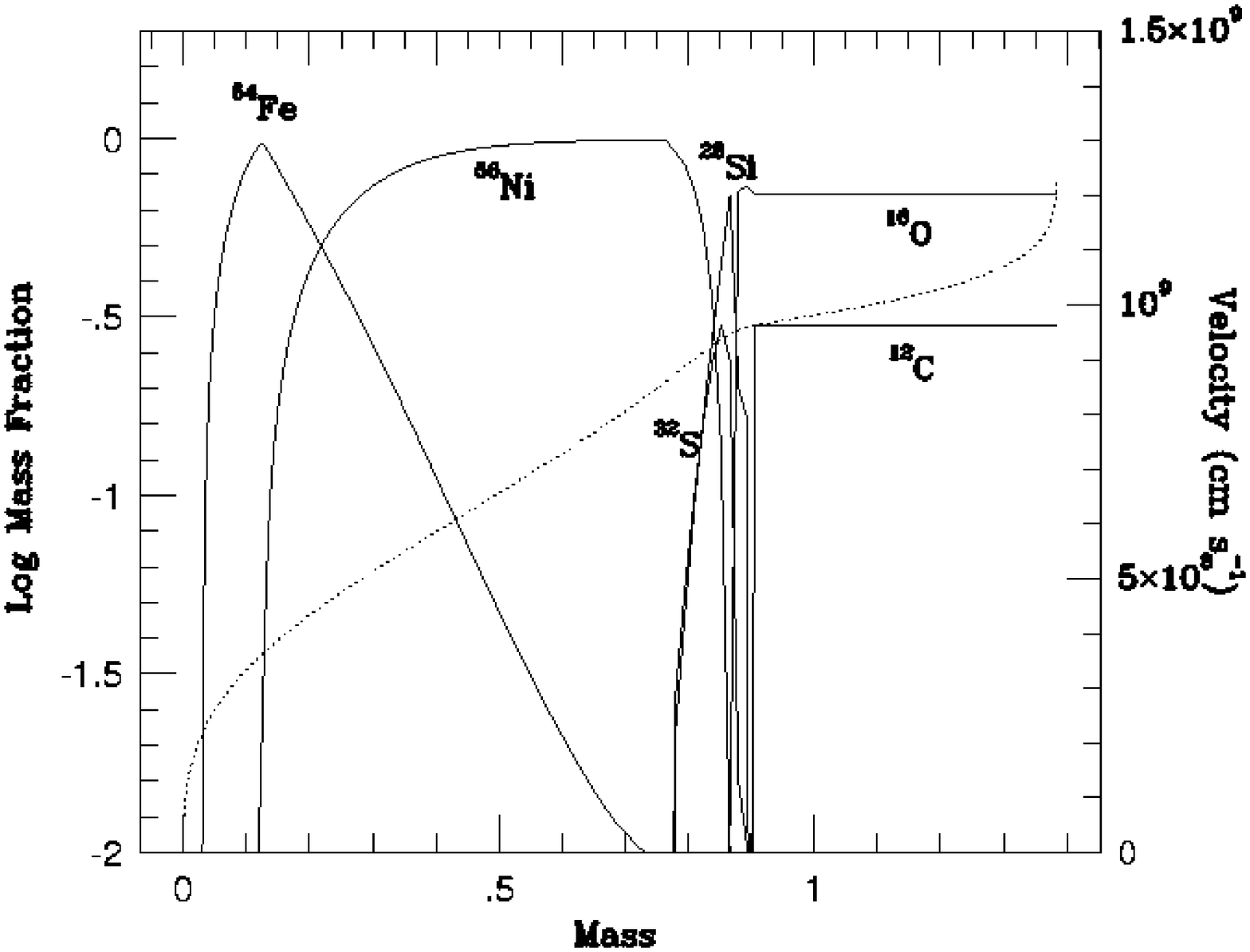}
\caption{\label{sharpw} Final velocity and composition in a simple one 
dimensional calculation of a deflagration in which the flame was
constrained to move at a speed (in the comoving frame) of $0.1 g_{\rm
eff} (t-t_o)$ with $t_o$ = 0.5 s. The velocity is sampled at t = 10 s
when it has taken on its coasting value. The final kinetic energy is
$8.5 \times 10^{50}$ erg. The mass of $^{56}$Ni is 0.57 M$_{\odot}$ \ and
there are 0.24 M$_{\odot}$ \ of other iron group isotopes.}
\end{figure}
A simple one dimensional calculation illustrating this possibility is
given in Fig.~(\ref{sharpw}). A Chandrasekhar Mass white dwarf was
ignited at its center and given an effective flame speed that, after
some time (0.5 s) during which the flame was assumed to move at a slow
speed (approximately laminar), was assumed to be given by
eq.~(\ref{vsw}). $g_{\rm eff}$ was calculated each time step at the flame
boundary. Calculations were done using the one-dimensional implicit
hydrodynamics code, Kepler (Weaver, Zimmerman, \& Woosley 1978), also
described in (\ref{mcrit}). This prescription assumes that the flame
moves with a radial speed given for an ensemble of RT-unstable bubbles
(Sharp 1984; \ref{rt}) and that turbulent processes behind the leading edge can
```digest'' all unburned fuel. The explosion is a healthy one, though
short of intermediate mass elements.

If for some reason the speed was greater, a stronger deflagration
would result. Effects that have been left out of the simple
Sharp-Wheeler speed (eq.~\ref{vsw}) are bubble growth by burning,
which causes even small bubbles to merge, the negative heat capacity
of a mixture of $^{56}$Ni 
and $\alpha$-particles (Garcia-Senz \& Woosley 1995), and gradients in
density and pressure.

\subsubsection{Active turbulent combustion?}
\label{act}
In section (\ref{tur}), we have neglected any feedback of burning itself on
the intensity and spectrum of turbulence. In particular, we assume
that the overall
propagation speed of the turbulent flame brush is limited by the
velocity of large scale RT-bubbles (equation \ref{vrt}). This need not
be true.  Owing to the limited range of observable scales, neither
simulation nor experiment presently rule out the possibility that
thermal expansion of burned material inside the flame brush
significantly increases
the strength of turbulent fluctuations on all turbulent scales.
The energy available from nuclear burning is
$\sim$10$^{18}$ erg g$^{-1}$, most of which goes into internal energy
and bulk expansion of the star. Only a tiny fraction of this energy,
$\sim$10$^{14}$ - 10$^{15}$ erg g$^{-1}$, is injected into turbulence
by the RT instability. Thus, the efficiency of any additional feedback
mechanism that converts nuclear energy to turbulence need not be
large. We therefore briefly discuss a possible
mechanism for the feedback of burning on turbulence and its
consequences for supernova explosions.

Thermal expansion behind the laminar flame is the main reason for the
LD-instability (section \ref{inst}). The growth of LD-perturbations in
the absence of external fluctuations is stopped by cusp formation
leading to a stable cellular flame structure. It is possible, however,
that once the cellular structure is disrupted by the turbulent cascade
from above, the production of specific volume contributes again to the
growth of velocity fluctuations. Although local intersections of flame
segments resembling cellular cusps still occur in the turbulent
regime, their random orientation inhibits the formation of any
self-stabilizing structure. Instead, it is likely that neighboring
expanding regions accelerate the medium in some
stochastic direction, thereby enhancing the intensity of velocity
fluctuations.

In a first approximation, we assume that this effect is most efficient on
the scale of the expanding regions themselves and therefore does not
significantly couple turbulent velocities on different scales. As a
consequence, only the fluctuation amplitudes would be affected, not their
spectrum. The simplest case would then be a constant and scale independent
growth rate, $\omega_{\rm tur}$, of turbulent velocity fluctuations on each
scale that depended only upon the expansion parameter $\gamma$,
\begin{equation}
\frac{d v(l)}{dt} = \omega_{\rm tur}(\gamma)\,v(l)\,\,,
\end{equation}
giving rise to exponential growth:
\begin{equation}
\label{expgr}
v(l,t) = v(l,0)\,e^{\omega_{\rm tur}(\gamma) t}\,\,.
\end{equation}
Since the most plausible functional dependence of $\omega_{\rm tur}$ on
$\gamma$ would be monotonously growing, this effect could readily
account for the required delay of flame acceleration for a delayed detonation.
Furthermore, the argumentation of section (\ref{tur}) concerning the transport
and burning time scales is still true, so that equation (\ref{utur}) still
holds on every scale above the Gibson length (which now, of course, decreases
according to the increasing strength of turbulence). More complicated, scale
dependent feedback mechanisms would also alter the spectrum of turbulent
velocities, giving rise to a fractal dimension of the flame surface different
from $D \approx \slantfrac{7}{3}$.

A recent proposal by Kerstein (1996) gives strong support to the idea
of active turbulent combustion. The author states that power-law
growth of the turbulent burning velocity (as opposed to exponential
growth expressed by eq.~\ref{expgr}) naturally occurs in the absence of
a stabilizing mechanism, either by self-acceleration or through a
statistical effect. However, the growth exponent is sensitive to the
underlying model of flame dynamics and thus remains undetermined.

As there is no natural limit for this growth in the incompressible
regime, the velocity of turbulent eddies on large scales would
eventually exceed that of rising RT-bubbles (equation \ref{vrt}). From
there on, the properties of the flame brush would be 
independent of the RT-instability. As the turbulent velocities become
close to sonic, compressibility effects grow more important and may
give rise to a delayed detonation (see section \ref{dd1}).
It must be noted that this self-enhancement effect can only occur
after the cellular structure has been disrupted above the Gibson scale by the
turbulent cascade from larger scales. This, together with the
increasing gas expansion as the star expands, may explain the delay in
the acceleration of the turbulent burning front necessary to account for
SN Ia nucleosynthesis. We emphasize again that the arguments given
here are speculation and need to be confirmed experimentally
or numerically. Three dimensional simulations of burning in a
pre-turbulized medium covering as many scales as reasonable are
needed.

\subsubsection{Initial conditions - multi-point ignition}
\label{sync}

As pointed out by Garcia-Senz and Woosley (1995) and demonstrated by
Niemeyer \ea (1996), the
efficiency of a white dwarf deflagration is sensitive to the manner in
which the runaway is
initiated. Multiple point ignition a few hundred km out from the
center gives more powerful explosions than one ignited at the stellar
center. This is because burning bubbles born off-center are more
buoyant at birth due to the higher gravity and can travel farther
before the star begins to disrupt.

It is presently uncertain how far this trend can be
continued. Obviously a white dwarf ignited simultaneously at at a
large number of points scattered uniformly throughout its mass would
have no difficulty exploding, but that would be artificial. The
condition of simultaneity requires synchronization of the burning in
regions, that at the time they run away, are out of communication. The
calculations of Garcia-Senz \& Woosley indicate that it would take an
extreme fine tuning of the temperature in order to get a convective
bubble in the phase immediately preceding runaway to float more than
$\sim$200 km. Still a large number of 10 km bubbles, say, could be
crowded on a sphere of this radius. If these floated a large distance
after burning before acquiring sufficient velocity to drive lateral
burning by the Kelvin Helmholtz instability and turbulence, the
effective burning rate could become quite large (picture a dandelion
gone to seed as the location of burned ash at this point). The
time-dependent competition between lateral and radial burning is thus
important. Three dimensional calculations to explore this possibility
would be interesting.

\subsection{Detonation}
The simplest form of explosion, in so far as the hydrodynamics is
concerned, would be detonation. This was the solution originally
proposed by Arnett (1969), but numerous variations have
been proposed since then that differ in where and when the transition from
subsonic to supersonic burning occurs. In order to make this
transition it is necessary that some volume sufficiently large to
maintain a detonation burn in a time short compared to that required
for sound crossing. Before discussing specific models, we begin by
deriving a grid of critical masses capable of igniting and sustaining
a detonation. We concentrate on densities between 10$^7$ and 10$^8$ g
cm$^{-3}$ because this is the approximate range required for a
transition to detonation that will provide ample intermediate mass
elements. 

\subsubsection{Critical masses for detonation}
\label{mcrit}

The general idea of a critical mass for detonation follows Blinnikov
\& Khokhlov (1986, 1987), Khokhlov (1991a), and Woosley (1990); see
also He \& Clavin (1994) for a recent analysis of critical conditions
for detonations. A
propagating detonation wave must exert sufficient over-pressure to
burn a region in a sound crossing time. If it does not, the shock will
degenerate into a pressure wave and damp. The shock temperature for
the self-consistent wave is given by the density and composition. This
temperature gives a time scale for burning which when multiplied by
the sound speed gives a rough estimate of the detonation wave
thickness. Unless this distance is very small compared to  
the size of the region where the phase velocity of burning is 
supersonic, the detonation will damp.

\begin{figure}
\plotone{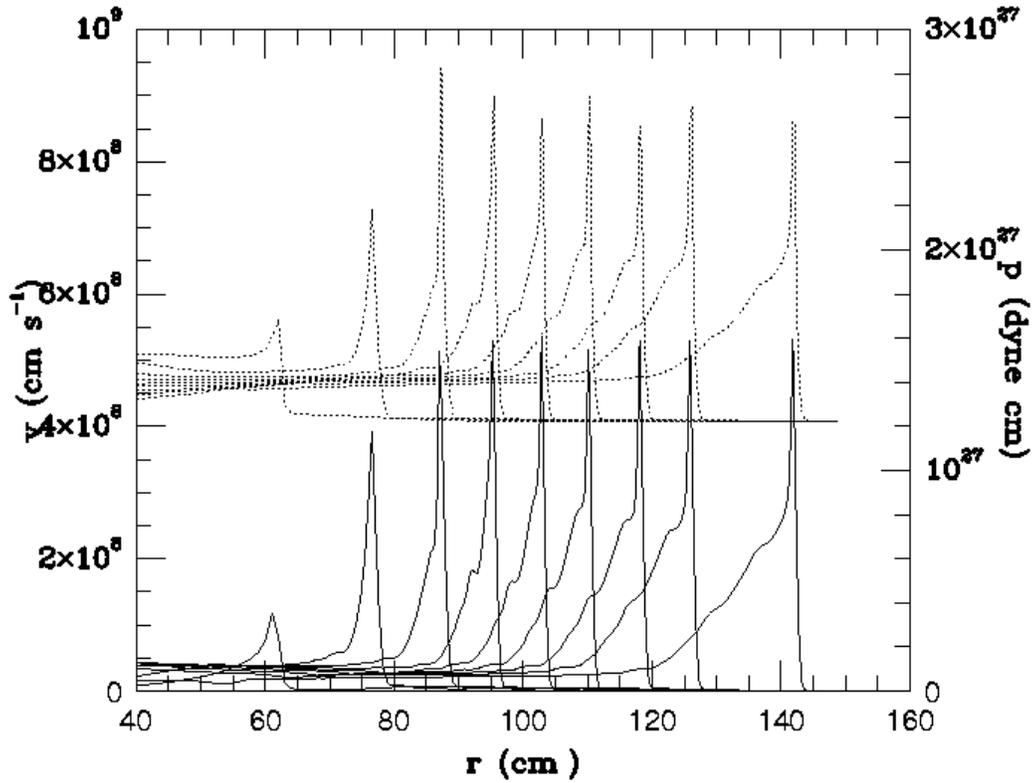}
\caption{\label{microdet} Microdetonation of a sphere of 50\% carbon 
and 50\% oxygen at $2 \times 10^9$ g cm$^{-3}$. The detonation was
intitated by a temperature gradient that gave a supersonic phase
velocity in the inner 70 cm (10$^{-18}$ M$_{\odot}$). The temperature fell
smoothly in the initial model to a constant background of $7 \times
10^8$ K at 87 cm ($2.7 \times 10^{-18}$ M$_{\odot}$). The density jump in
the steady state shock was 1.5. The times sampled in the figure are 
0.66, 1.9, 2.6, 3.2, 3.8, 4.3, 4.8, 5.4, and 6.5 $\times
10^{-8}$ seconds. Velocity is the solid curve; pressure is the dashed
curve.}
\end{figure}

We have determined empirically a series of critical masses using the
one-dimensional hydrodynamics code, Kepler (Weaver, Zimmerman, \&
Woosley 1978), which has nuclear physics and an equation of state
appropriate for the problem. The nuclear network employed contained 19
isotopes from hydrogen to $^{56}$Ni. Radiation transport was turned
off and only the quadratic artificial viscosity term employed. Spheres
were constructed of 100 zones with the mass of each zone smoothly
increasing outwards (the first 30 zones had constant mass, the other
70 gradually increased logarithmically to a value 100 times larger
than the central zone in zone 100). For the calculations at $\rho \le
10^8$ g cm$^{-3}$, a linear (with respect to interior mass)
temperature gradient was superimposed on the inner 28 zones with a
central value of 3.2 $\times 10^9$ K falling to $4 \times 10^8$ K in
zone 28. The rest of the zones also had a temperature of $4 \times
10^8$ K. This temperature gradient was such as to give a well resolved
supersonic phase velocity for the burning in the inner zones for all
spheres considered. The density was taken to be approximately constant
in all zones and a boundary pressure applied to the sphere equal to
that in the outer zones. For the $\rho = 2 \times 10^9$ g cm$^{-3}$
case, the central temperature was assumed to be $2.8 \times 10^9$ K
declining smoothly over 55 zones to $7 \times 10^8$ K where it
remained constant for another 145 zones. A sample calculation is given
in Fig.~(\ref{microdet}).  

Burning was then turned on and the subsequent evolution followed for
about 4000 times steps in each case (20,000 in the case of $2 \times
10^9$ g cm$^{-3}$). Several dozen such spheres were modeled. A sampling
of results is given in Table 1. In each case the critical mass is the
lowest value for which detonation was achieved and maintained. The
value given is the mass interior to zone 20, i.e., most of the
material upon which the high burning temperature was imposed. In each
case a calculation using 10 times less mass was also carried out and
gave a failure.  The length scale is thus resolved to about a factor
of two. Because the results depend so sensitively upon the
composition, it was not worthwhile to attempt greater resolution.

For the case of 50\% carbon and 50\% oxygen our results agree well
(within a factor of 10) with those of Arnett \& Livne (1994b). However,
we also explored the sensitivity to composition and found the mass
fraction of carbon to be very important.  At a density of $3 \times
10^7$ g cm$^{-3}$, for example, the critical mass is 5 orders of
magnitude smaller for pure carbon than for a mixture of 50\% C and O.
The larger Q-value for the burning gives a higher burning temperature
in the detonation and the reactions are very temperature sensitive. On
the other hand, the mass is five orders of magnitude {\sl larger} if
the fuel is 40\% carbon and 60\% silicon, and for 35\% carbon and 65\%
silicon the critical mass is larger than the star. We calculated
mixtures of silicon and carbon because, later, we shall consider the
possibility of igniting a detonation in a region where ash and fuel
have mixed. At these densities, carbon and oxygen burn to intermediate
mass elements of which silicon is more representative than
nickel. Apparently, for each density, there is a critical mass of
contamination by ash that makes detonation impossible. But if the ash
remains a small component, the required critical masses are quite
small.

While the calculations were carried out for thoroughly mixed combinations,
similar restrictions would apply to a region of thinly laminated, but unmixed
fuel and ash. The critical quantity is the overpressure produced in a region
large compared to the mixing scale. Since the ashes have long since yielded
their overpressure to sustaining the subsonic expansion of the star, only the
contribution of the fuel is important. If the mass fraction of fuel in the
macroscopic region is low, the critical mass is large.

These masses are for isolated hot spots surrounded by cold material.
If the runaway occurs in an extended region that is itself close to
burning on a sound crossing time, they could be smaller.

\subsubsection{Prompt detonation}
\label{prompt}
We now discuss the possibilities for detonation in the approximate
time order in which the detonation appears.

First, there is the possibility of a prompt detonation (Arnett
1969).  Critical masses for prompt detonation have been
calculated by Blinnikov \& Khokhlov (1986, 1987) and are very small
(see also Table 1 for $\rho = 2 \times 10^9$ g cm$^{-3}$), so there
are an enormous number ($\sim$10$^{18}$) of possibilities for
detonation in the inner 100 km of the white dwarf that runs away. At
issue, however, is the high degree of isothermality required to
produce a supersonic phase velocity.

Woosley(1990) derives the condition that the phase velocity of nuclear
burning be sonic in the presence of a temperature gradient $dT/dt$:
\begin{equation}
\label{gradt}
\left(\frac{d T}{d r}\right)_{\rm crit} \ = \ 0.3 \, T_9^{21.2} \,
\rho_9^{3.05} \ {\rm K \ cm^{-1}}.
\end{equation}
Convection ceases to be efficient in transporting energy at about $T_9
\ = \ 0.7$, when the convective cycle time equals the nuclear burning
time, and this is a convenient point at which to evaluate the
isothermal condition. The above equation gives a critical gradient,
$dT/dr \sim$ 0.001 K cm$^{-1}$.  Conduction will tend to smooth out
temperature fluctuations on small scales. For $\rho = 2 \times
10^9$ g cm$^{-3}$ and T$_9$ = 0.7, the conductive opacity is about
10$^{-5}$ cm$^2$ g$^{-1}$ and the conductive length scale for 1 s,
$\sim$10 m.  The time scale to increase nuclear energy generation at
$T_9$ = 0.7 is $\sim$10 s. Interestingly the critical mass
for detonation (Table 1) is also a few meters in radius, so perhaps an
approximately isothermal region of this size is not unreasonable.

However, for detonation within a single critical mass of 10$^{-18}$
M$_{\odot}$ \ (Table 1), the temperature could only vary from $T_9$ = 0.7 by
about one part in 10$^{10}$ (if the conditions at $T_9 \approx 0.7$
evolved without outside influence to runaway)! For larger length
scales, up to 10$^7$ cm, the restriction is less severe, but still very
constraining, about one part in 10$^5$. Further, this thermal
condition must be maintained for an appreciable fraction of the
nuclear time scale at $T_9$ = 0.7,
\begin{equation}
\label{taunuc}
\tau_{\rm nuc} \ = \ 0.15 T_9^{-20.2} \rho_9^{-3.05} \ \approx \ {\rm
20 \ s}
\end{equation}

A prompt detonation becomes more likely if an only approximately
isothermal region exists at higher temperature (eq.~\ref{gradt}). For
example, consider the conditions after the runaway has progressed and
the hottest temperature anywhere in the core is $T_9$ = 1.2. If this
point is surrounded by say, 10 km of material in which the temperature
is between $T_9$ = 1.2 and 1.1, a detonation will occur. Similarly a
point having $T_9$ = 1.5 must be surrounded by 10$^{4}$ cm of material
between $T_9$ = 1.4 and 1.5, etc. However, the nuclear time scale at
these points are $5 \times 10^{-4}$ s and $5 \times 10^{-6}$ s
respectively, and there is no efficient energy transport mechanism
that operates on such short times.

The actual thermal structure is not known with such
precision. Small fluctuations will be amplified by the temperature
sensitivity of the nuclear burning. Turbulence, energized by
convection, has two effects. On the one hand, it acts to smooth out
inhomogeneities. But, to the extent that the
energy density in turbulence varies even slightly from place to
place, different amounts of heat will be dissipated at the (very small)
Kolmogorov scale. The convective velocity at $T_9 = 0.7$ is
approximately 30 km s$^{-1}$.   The energy density in this turbulence is
about 10$^{12}$ - 10$^{13}$ erg g$^{-1}$; the heat capacity of the
gas, $\sim$10$^7$ erg g$^{-1}$ K$^{-1}$. The time scale for generating
this energy is on the order of the convective turnover time 10 - 100
s. If 10\% variations occurred in the dissipation rate in two regions,
the temperature variation would be of order 10$^5$ K. 
Though this does not prove that a prompt detonation is impossible, it
suggests that it is much more difficult than the large number of critical
masses implies.

The chief evidence against prompt detonation is not so much the
physics of igniting the detonation, as the fact that the models do not
agree with observations. They produce too much $^{54}$Fe with respect
to $^{56}$Fe and no intermediate mass elements (e.g., Woosley 1990).
The detonation wave may be unstable at high densities ($\rho
\gtrsim 2 \times 10^7$ g cm$^{-3}$; Khokhlov 1993b), but the
instabilities predicted act on scales of order one cm and less. This
would have no effect on the nucleosynthesis since even a laminar
flame would burn through the residual fuel in less than an expansion
time. If an instability were found that left behind regions of
unburned material with the size of order 100
km, these regions would not be completely burned by conduction
in the explosion. The model would resemble the deflagration ignited at
many points throughout its volume discussed in the previous
section. To our knowledge, no such instability has been proposed.

\subsubsection{Atmospheric detonation}
\label{astmos}

The first models in which detonations were actually calculated, rather
than assumed (Nomoto, Thielemann, \& Yokoi 1984; Woosley \& Weaver
1986) occurred because a subsonic deflagration produced sufficiently
strong pressure waves that their accumulation ahead of
the flame led to compression adequate to ignite nuclear burning on a
sonic time. This always occurred near the surface of the star
in the steepening density gradient around 1.0 - 1.3
M$_{\odot}$. Such explosions made little intermediate mass elements, but did
produce very high velocity iron synthesized as $^{56}$Ni near the
surface. The high velocities may be necessary to explain some
supernova observations, especially for SN 1991T (Yamaoka et al. 1992),
but the small synthesis of Si - Ca remains a problem for the typical
SN Ia. The reason so little silicon is made in these models is that
the flame must move an appreciable fraction of the sound speed in
order to steepen into a shock in the outer layers. Such rapid burning
consumes the white dwarf before it has had time to expand and is still
at high density.

To produce intermediate mass elements, it is helpful if the supernovae
first burns relatively slowly and expands and then makes a transition to
a detonation at a density $\sim$10$^7$ g cm$^{-1}$. At such a low
density detonation does not proceed to iron but to silicon. This
defines the ``delayed detonation'' model of Khokhlov (1991abc) and
Woosley \& Weaver (1994).

\subsubsection{Delayed detonation}
\label{dd1}
First we consider the model of Khokhlov who attributes the transition
to detonation to temperature fluctuations. These fluctuations
arise from ``non-uniform preheating of the gas'' (or perhaps
turbulence, though he emphasizes the former). Khokhlov points out that
a spontaneous transition to detonation is frequently observed in
terrestrial experiments (e.g., Lee \& Moen 1980) involving turbulent
flames. The underlying idea is similar to that discussed in
(\ref{prompt}). A single small region, but one larger than the critical
mass, burns faster than sound because of an anomalously high
temperature and shallow temperature gradient. Once a detonation is
initiated, even a small one, it propagates through the rest of the
star.

However, such large localized temperature fluctuations are unlikely. So
long as the flame retains a well defined surface (i.e., operates in
the a laminar or ``flamelet" regime; \ref{tur}), high fuel temperatures
exist only in the thin interface separating fuel and ash. Burning
in this interface is very subsonic (defining, in fact, the laminar
speed) and the mass in a flame thickness, much less than a critical
mass. Previously existing, isolated temperature fluctuations in the
carbon fuel will tend to be damped and quenched by the expansion. By
very careful tuning, one might arrange to have a disconnected region
run away at a late time (burning would occur at a rate that, for a
long time, precisely balanced adiabatic losses as in Garcia-Senz \&
Woosley 1995). Additional tuning could give this region such a shallow
temperature gradient that it burned supersonically. Not only does this
seem unlikely, but the runaway would be much more likely to occur at
an earlier time and there would be nothing special to delay the
burning until the density declined below $3 \times 10^7$ g cm$^{-3}$,
as is necessary if intermediate mass elements are to be produced.

Temperature fluctuations might also be built up by the collision of sound
waves or blobs of matter moving in random directions in the unburned carbon,
but the energy density of these collisions is small. Typical velocity shears
are 100 km s$^{-1}$ implying an energy density of $\sim 10^{14}$ erg
g$^{-1}$. At a density of $3 \times 10^7$ g cm$^{-3}$, an internal energy
addition of about 10$^{17}$ erg g$^{-1}$ is required to raise the temperature
above 2.0 billion K where burning on a hydrodynamic time scale can occur. We
conclude that a delayed detonation initiated by temperature fluctuations is
unlikely.

A different sort of delayed detonation was discussed by Woosley \&
Weaver (1994; in lectures presented at Les Houches in 1990). This kind
of model resembles more the atmospheric detonations of
section (\ref{astmos}), but there is an attempt to ``sculpt'' the
flame speed so as 
to provoke a detonation in a star with initially slow burning. In
outcome the model is the same as that of Khokhlov, but the detonation
occurs because of an accumulation of overpressure in a macroscopic
region surrounding a topologically complex flame surface. Geometry
plays the key role, not temperature fluctuations or gradients, and 
it makes sense to cast the model in terms of the fractal
dimension of the flame. Some representative conditions illustrate the
idea. At $ 3 \times 10^7$ g cm$^{-3}$, the critical mass for
detonation is around 10$^{-14}$ M$_{\odot}$ \ (50 meters for 50\% carbon;
Table 1).  The turbulent flame brush (section \ref{tur}) might be up
to 200 km thick (i.e., it contains many critical 
masses; Niemeyer \ea 1996), the laminar speed of the flame, about 0.5 km
s$^{-1}$, the sound speed near 5000 km s$^{-1}$, and the smallest
unstable wavelength, about a centimeter. The 200 km thick region will
then burn supersonically if the fractal dimension in that region
exceeds 2.6.

There are several difficulties here. First, the fractal dimension
associated with a turbulent flame is well pegged to 2.3 - 2.36 based
both upon simple scaling relations (Kerstein 1988, 1991) and experiment
(e.g., Mantzaras \ea 
1989, North \& Santavicca 1990, Haslam \& Ronney 1995). A value of 2.6
would require 
non-standard assumptions (though Mandelbrot, 1983, offers arguments
that the fractal dimension of isoscalar surfaces in turbulence should
be 2.5 to 2.66). It should be recognized however that $D =
\slantfrac{7}{3}$ is a statistical average. There may be regions
having larger dimension and others with smaller ones.

Why would this detonation occur preferentially at late times? At
earlier times, the Gibson length is much larger and the structure of
the flame brush correspondingly coarser. At 10$^9$ g cm$^{-3}$, for
example, the laminar flame speed is about 50 km s$^{-1}$. A flame
brush of 200 km (though none would exist yet at this early time) would
have a sound crossing time of about 0.02 s and the spacing between
burning regions for supersonic effective burning would be 1 km. The
Gibson length at these conditions is larger than 1 km; so too is the
minimum unstable Rayleigh Taylor wavelength (Timmes \& Woosley
1992). It would be hard to prepare such a layered structure without
burning all the fuel in the process. Detonation would be difficult.

A second condition required for a delayed detonation of this type is
that the mass fraction of ash in the flame brush not be
high. At $3 \times 10^7$ g cm$^{-3}$, Table 1 suggests that detonation would
not occur if the ash comprised more than 35\% of the mass of the flame brush.

We conclude that a delayed detonation of the kind proposed by Woosley
\& Weaver (1994) is possible, but its actual occurrence improbable.
Whether the requisite fine structure can be set up is unknown. 

\subsubsection{Detonation in the distributed regime}
\label{dist}
Instead of mixing and wrinkling a thin flame sheet, turbulence, in the
extreme, might mix heat from the ashes into cold unburned
fuel. Rather than accumulate a critical surface area within a volume,
the star might instead gradually build a critical temperature in a
volume of fuel larger than in Table 1, as Khokhlov's model (1991abc)
requires. This sounds straightforward, but can be difficult to
arrange.  Indeed, for densities above about $3 \times 10^7$ g
cm$^{-3}$, without pulsation, it is impossible.

Above this density, the microphysics of the nuclear burning and
electron conduction can be decoupled from all the instabilities we
have discussed.  Viewed on a sufficiently small scale (i.e., the
conductive flame thickness), the flame is smooth and stable
(Fig.~\ref{ul}). Since the thickness of the flame is smaller than any
critical mass in Table 1, no detonation can occur.

However, as the density drops below about $3 \times 10^7$ g cm$^{-3}$
(Fig. \ref{trans}) nuclear burning proceeds at a slower rate and the
flame becomes thicker.  For a mixture of 50\% carbon and 50\% oxygen, Timmes \&
Woosley (1992) find a flame thickness of 0.5 cm at $5 \times 10^7$ g
cm$^{-3}$ and 4 cm for $1 \times 10^7$ g cm$^{-3}$. At the same time,
the scale of the smallest turbulent eddies that can turn over without
burning (the Gibson scale) grows progressively smaller. For a
turbulent energy density of 10$^{14}$ erg g$^{-1}$ at 10$^6$ cm
(Niemeyer \& Hillebrandt 1995b), the Gibson scale is 0.2 cm and
10$^{-4}$ cm, respectively at 5 and $1 \times 10^7$ g
cm$^{-3}$. Somewhere between these two densities a transition to a
different kind of burning must occur.

Under these conditions it no longer makes sense to speak of a flame
propagated solely by electronic conduction. The burning is smeared out
by turbulence; one enters the distributed regime of
Fig.~(\ref{trans}). Heat can then be extracted from burning regions
and transported to fuel.  To make the transition to detonation a
region of size larger than the critical mass (Table 1) must assume a
temperature gradient shallower than (\ref{gradt}) with a peak
temperature such that the nuclear burning time is much shorter than
the stellar expansion time. We now illustrate, with specific conditions,
how this might occur.

First, a portion of the flame brush moves into the distributed
region. We take a density $3 \times 10^7$ g cm$^{-3}$ as
representative. The turbulent flame brush is, at this point,
a fine grained mixture of discrete 
phases - fuel and ash, fine grained, but still separated. Now, in
places, the burning begins to go out as turbulence penetrates the flame sheet,
homogenizes the composition, and reduces the temperature to some low
mean value where burning is very slow. Recall the physics of the
laminar flame (Timmes \& Woosley 1992). There is a critical
temperature in the flame where conduction balances nuclear energy
generation. The nuclear time scale at this temperature times the
laminar flame speed equals its thickness. As the flame enters the
distributed region, turbulent eddies with a size comparable to the
flame thickness ($\sim$ 1 cm) become fast enough to carry away heated
fuel before it can burn. One might say that turbulence, for the
first time, ``enters'' the reaction region
and disperses it to larger length scales where, again, the burning
time equals the (turbulent) transport time. For still lower densities
and temperatures, and thus lower burning rates, the same level of
turbulence is able to distribute the reaction zone on even larger scales.

This leads to regions where the burning is temporarily quenched.  As
the overall density declines, these regions grow in size and, if the
star becomes unbound, eventually encompass all the flame brush. These
regions of suppressed burning are still coupled by turbulence,
however. An important time scale to keep in mind is the characteristic
turbulent time scale for a critical mass (Table 1). We shall assume,
and justify later, that the fraction of ash in the mixture is
small. Taking carbon equals oxygen equals 50\% by mass, the critical
mass is $\sim 10^{-14}$ M$_{\odot}$, and its size 50 m. Its turbulent turnover
time is then $\tau_{\rm crit} \sim 10^{-3}$ s (Fig. \ref{ul}). It is
important that this is very much less than the hydrodynamical time,
$\tau_{\rm dyn} = 446/\rho^{1/2} \sim$ 0.1 s, for the star to expand.

If the mixture of fuel and ash burns faster than $\tau_{\rm crit}$,
mixing will be incomplete and no detonation is possible. This requires
that the temperature of the mixture be less than $T_9 = 2.2$
(\ref{taunuc}), cooler if one considers larger regions containing many
critical masses. On the other hand, the mixture must be able to resume
rapid burning during the hydrodynamic time scale and that requires a
temperature {\sl larger} than $T_9 = 1.7$. For intermediate
temperatures of the mixture, a detonation is possible. The temperature
in the ashes of a typical deflagration when the density is $3 \times
10^7$ g cm$^{-3}$ is $T_9$ = 4 to 5. The electronic heat capacity
still dominates, so $C_{\rm v} \propto$ T. Therefore mixing equal
masses of fuel and ash gives a mixture with temperature $T_{\rm
ash}/2^{1/2}$, with little dependence on the temperature in the cold
fuel. If we want the mixture to have a temperature of $T_9$ = 2, for
example, then each part of ash must be diluted with 5 parts of fuel if
the unmixed ash has temperature $T_9$ = 5 and 3 parts fuel if it has
temperature $T_9$ = 4. This justifies the use of the (ash-free)
critical mass chosen above. The actual dilution factor must ultimately
come from a numerical study that we are unable to do right now. If it
is too large or too small the model fails. 

A delay follows during which the temperature rises and the mixture
runs away. For a detonation to occur, a region larger than the
critical mass must have a nearly isothermal temperature distribution,
(\ref{gradt}), about 20 K cm$^{-1}$ if we continue to use our (very
approximate) representative value of $T_9$ = 2. This means is that
once the mixture resumes burning, there can be no
large region cooler than this peak value minus the temperature gradient
times the size of the region (10$^5$ K for 50 m). The region can, of
course, be larger than the critical mass and the condition on the
temperature fluctuations less stringent, but larger regions will
have longer turbulent time scales and it will be harder to mix them
without the fuel already burning.

As in the case of prompt detonation (\ref{prompt}), one has many
opportunities for an improbable event, so the outcome is
uncertain. However, because of the higher characteristic temperature
($T_9 \approx 2$ rather that $T_9 \approx 0.7$), the isothermal
condition is not nearly so stringent, and with as many as $\sim
10^{12}$ critical masses in the flame brush, a detonation does not
seem so unlikely.  The congruence of the density where distributed
burning can begin with that required for a detonation to make
appreciable intermediate mass elements is also particularly
encouraging for this model - another key difference with the prompt
detonation model.

\subsubsection{Pulsational Detonation}
\label{pulsdet}

If inadequate fuel burns to disrupt the white dwarf on the first
pulse, contraction will cause rekindled combustion and one or more
pulsations will ensue. The burning conditions after each pulse will
differ appreciably from the one before. If the pulse does not go below
$\sim 3 \times 10^7$ g cm$^{-3}$, the boundary between ash and fuel
remains intact, but the flame surface becomes dispersed throughout a
greater fraction of the mass and more convoluted as well. If the pulse
goes below $\sim3 \times 10^7$ g cm$^{-3}$, turbulence will mix both
the composition and the heat across the interface.  During the
recompression, a large region of shared energy (or of dispersed flame)
will reignite giving rise once more to the possibility of detonation,
or at least very rapid combustion.

Models of this sort have been studied by Khokhlov (1991b); Khokhlov,
M\"uller, and H\"oflich (1993); H\"oflich, Khokhlov, and Wheeler
(1995) and, in two dimensions, by Arnett \& Livne (1994ab) following
early pioneering work on pulsational deflagration by Nomoto, Sugimoto,
\& Neo (1976) and Ivanova \ea (1974). More recently Woosley (1996) has
demonstrated, using mixing length convection theory and a fractal
flame in a one-dimensional model where pulsation was explicitly
calculated, several possible outcomes including either detonation (the
pulsational analog to \ref{dd1}) or an accelerated deflagration.
However, no previous calculation has properly considered the critical
role of turbulence.  Without turbulence, the surface topology is not
made much more complex during the pulse, nor is heat appreciably shared
between fuel and ash. The heated region is instead confined to a
narrow layer on the surface of a conductive flame. This is the
definition of the flame thickness and, for a monotonically expanding
supernova, this thickness is, at all times, thinner than the critical
mass for detonation. It is possible, in principle,
for a very large amplitude pulsation to lead to such a thick
conductive flame that, during recontraction, the heated region
encompasses a critical mass, an approach taken, for example, by Arnett
\& Livne (1994b).  But for the same conditions, turbulence would
dominate the heat transport.

Observationally, the pulsational detonation model, or at least the
single large pulse version, has a difficulty. If a large amplitude
pulse is necessary for the initiation of the detonation, one would
expect {\sl some} white dwarfs to explode weakly without pulsing at
all. Why should all initial pulses fall just short of producing an
explosion? These supernovae with their faint broad light curves have,
at the present time, no observational counterpart. Perhaps they await
discovery. Or perhaps, the explosion proceeds through a series of low
energy pulses, only the last of which is always adequate to provoke
detonation. In this latter case, the density $3 \times 10^7$ g
cm$^{-3}$ is critical. Pulsations that do not go below this value
would not lead to greatly accelerated burning because their flames
would remain, on the small scale, laminar and dominated by conduction.
The first pulsation to go below this density would experience
appreciable mixing and heat sharing and greatly accelerated burning,
perhaps detonation, on the next pulse. Burning on a hydrodynamic time
scale would also be guaranteed since burning would be responsible for
halting the recompression.

What actually happens will not be known until realistic three
dimensional models (with realistic ignition conditions; \ref{sync})
have been calculated, both to show the failure of the first pulse and
the degree of turbulent mixing during subsequent pulses.
\section{CONCLUSIONS}
\label{conc}

Our first conclusion, perhaps not a very reassuring one to the
observers, is that there are a lot of models for how a white dwarf
explodes, none of which can be definitely excluded. However, our
analysis suggests some models are more easily realized than others and
calculations to clarify the situation.

The simplest model, plain carbon deflagration, is an area where
progress can and needs to be made. Currently one does not know, in the
common case, whether the burning front, with all its instabilities, is
able to unbind the star without an intervening pulse, which would
exclude the pulsational detonation model, or not. If a radial front
moves at the Sharp-Wheeler speed (\ref{vsw}), the star becomes
unbound, but this assumes efficient combustion by turbulence within
the RT mixing layer (\ref{simdefl}). The radial velocity of the bubble
front is set by 
the non-linear Rayleigh-Taylor instability, which should advance at
nearly the Sharp-Wheeler speed (0.1 $g_{\rm eff} t$). Behind the
leading edge of the bubble front, burning is enhanced by turbulence
generated by the Kelvin-Helmholtz instability in shear flows bounding
the RT unstable blobs.  It is not clear that fuel consumption
proceeds at the same speed as the bubble front moves into the fuel. In
other words, a stationary turbulent flame brush may never be 
established during the explosion, in which case the burning
speed remains smaller than the Sharp-Wheeler speed. High resolution three
dimensional calculations should ultimately clarify the issue. If the
two speeds are comparable, an explosion seems likely, if only a mild one. The
success of the first pulse is also very dependent upon how the runaway
is initiated in the star (\ref{simdefl},\ref{sync}).

If the simple deflagration succeeds then, in order to agree with
observations, it must have a higher speed than current
multi-dimensional calculations suggest. We have discussed two possible
ways of speeding it up - active turbulent combustion (\ref{act}) that
increases the turbulence intensity by thermal expansion within the
flame brush, and extreme multi-point ignition (\ref{sync}). According
to recent results for the scaling behavior of unconfined turbulent
flames (Kerstein 1996), active turbulent combustion is a very
promising field for future investigations, but its influence on the
explosion mechanism remains speculative. 

Delayed detonations without an intervening pulsation are also not
excluded, but those in the current literature have problems
(\ref{dd1}). The temperature fluctuations required to induce
detonation by a ``spontaneous burning'' (Khokhlov 1991abc) are
unlikely and the large fractal dimension and small minimum wavelength
used by Woosley \& Weaver (1994) are inconsistent with current views
regarding turbulence. Indeed, in the current view, an effective
burning speed, $u_{\rm tur}$,  faster than the fastest turbulent
motion, $v(L)$, (which occurs
on the largest scale $L$) is impossible (section \ref{tur}). This form of
delayed detonation model survives for the time being, because the
Kolmogorov mean field description of turbulence may not fully describe
our time dependent situation. The relation $u_{\rm tur} \approx v(L)$ is an
average. There may be appreciable fluctuations from this mean in
isolated small regions larger than a critical mass (we are currently
investigating this). Active turbulent combustion (\ref{act})
might enhance the fuel consumption rate sufficiently. Another
possibility is a burning geometry that favored only radial growth
early on, but rapid non-radial combustion at late times.

A more likely possibility, which we are suggesting here in the astrophysical
context for the first time, is a transition to detonation as the star
expands, its density declines, and it enters the regime of distributed
burning (Fig. \ref{trans}; \ref{dist}). This kind of explosion has
several appealing characteristics. First, the transition to detonation
occurs as a deflagration is dying. This naturally leads to the desired
pre-expansion of the star and rapid burning at late times. In
particular, abundant intermediate mass elements can be
synthesized at densities below the transition density, $\sim 10^7$ g
cm$^{-3}$.  If the detonation ignites, it will naturally give a very
energetic explosion, currently a problem for the deflagrations.  Most
importantly it does these things using credible, definite physics,
which, though uncertain, can be tested by numerical modelling. At this
point its greatest uncertainties are the actual dilution factor for
fuel and ash (a value of one part ash to several parts fuel is
optimal) and whether the necessary isothermal conditions can be set up
in the mixed medium at $T_9 \approx 2$. {\sl For these reasons, this
is the model that that we favor at the present time.} The
observational consequences of this form of detonation should be the
same as for other forms of (parameterized) delayed detonation models. 

If the detonation fails to catch on the first try, and if simple
deflagration fails, then pulsational detonation (\ref{pulsdet}) offers
an appealing alternative, but not one without its own puzzles. The
deflagration must fail after having burned an appreciable
mass so that the dwarf experiences a large amplitude oscillation. 
This requires some tuning so that the star does not become unbound and
yet still expands enough to mix. Why do we see no supernovae that just
barely exploded? Here again distributed burning may help. The
detonation only lights once the pulsation has sufficient amplitude to
go to 10$^7$ g cm$^{-3}$ at which point efficient mixing occurs. The
star could approch this limiting density by one or more pulsations. 

Because fundamental physics does not yet preclude several
qualitatively different outcomes from very similar starting points, it
is possible that they all happen to some extent. Even within a single
class of model there is room for considerable variation owing to
different ignition conditions, variable ratios of carbon to oxygen
(which set the underlying conductive speed as well as affect the
critical detonation mass), and the uncertain transition to detonation.
Such diversity may be necessary to understand such
distinctively different Type Ia supernovae as SN 1991bg and SN 1991T.

\acknowledgements
This work has been supported by the National Science Foundation (NSF
91 15367) and the NASA Theory Program (NAGW 2525 and NAGS-2843) and,
in Germany by a DAAD HSP II/AUFE fellowship (JCN) and a Humboldt Award
(SEW). We thank Sergei Blinnikov for a careful reading of the manuscript and 
useful comments, and Wolfgang Hillebrandt for informative discussions. We are
particularly indebted to Alan Kerstein of Sandia Laboratory for
several illuminating discussions regarding turbulent chemical
combustion.

\clearpage
 
\begin{deluxetable}{rcccccccc}

\tablecaption{Critical Masses \label{tb1}}

\tablehead{
\colhead{$\rho$} & \colhead{$^{12}$C} & \colhead{$^{16}$O} &
\colhead{$^{28}$Si}   & \colhead{$^{56}$Ni} & \colhead{Mass} &
\colhead{Radius}    & \colhead{T$_{9burn}$}  &
\colhead{$\tau$\tablenotemark{a}} \nl 
& & & & & \colhead{[M$_\odot$]} & & \colhead{[10$^9$ K]}  & \colhead{[s]} 
} 

\startdata
10$^7$ & 0.5 & 0.5 & 0 & 0 & 10$^{-10}$ & 2 km & 3.5 & $4 \times 10^{-4}$
\nl
 & 0.5 & 0 & 0.5 & 0 & 10$^{-8}$ & 10 km & 3.3 & $2 \times 10^{-3}$ \nl
\nl
$3 \times 10^7$ & 1.0 & 0 & 0 & 0 & 10$^{-19}$ & 1 m & 5.5 & $2 \times
10^{-7}$ \nl
 & 0.5 & 0.5 & 0 & 0 & 10$^{-14}$ & 50 m & 5.2 & $1 \times 10^{-5}$ \nl
 & 0.5 & 0 & 0.5 & 0 & 10$^{-10}$ & 1 km & 4.3 & $2 \times 10^{-4}$ \nl
 & 0.5 & 0 & 0 & 0.5 & 10$^{-10}$ & 1 km & 4.1 & $2 \times 10^{-4}$ \nl
 & 0.4 & 0 & 0.6 & 0 & 10$^{-5}$ & 50 km & 3.9 & $1 \times 10^{-2}$ \nl
 & 0.25 & 0.25 & 0.5 & 0 & 10$^{-4}$ & 100 km & 4.0 & $2 \times 10^{-1}$ \nl
 & 0.35 & 0 & 0.65 & 0 & $\gtrsim 10^{-1}$ & - & - & - \nl
\nl
10$^8$ & 1 & 0 & 0 & 0 & 10$^{-20}$ & 40 cm & 6.5 & $1 \times 10^{-7}$ \nl
 & 0.5 & 0.5 & 0 & 0 & 10$^{-18}$ & 2 m & 6.2 & $4 \times 10^{-7}$ \nl
 & 0.5 & 0 & 0.5 & 0 & 10$^{-12}$ & 150 m & 5.5 & $3 \times 10^{-5}$ \nl
\nl
$2 \times 10^9$ & 0.5 & 0.5 & 0 & 0 & 10$^{-18}$ & 70 cm & 9.7 & $7
\times 10^{-8}$ \nl 
\enddata
\tablenotetext{a}{Sonic crossing time assuming a sound speed of 5000
km s$^{-1}$; (10,000 at $2 \times 10^9$)} 
\end{deluxetable}

\end{document}